\documentclass[preprint,aip,preprintnumbers,amsmath,amssymb]{revtex4-1}

\usepackage{graphicx}
\usepackage{dcolumn}
\usepackage{bm}
\usepackage{amsmath}

\newcommand{\sinc}{{\rm sinc}}
\newcommand{\rect}{{\rm rect}}

\begin{document}


\title{
Sinc-based method for an efficient solution in the direct space of
quantum wave equations with periodic boundary conditions}

\author{Paolo Marconcini} \email[Corresponding author: ]
{paolo.marconcini@iet.unipi.it}
\author{Demetrio Logoteta}
\author{Massimo Macucci}
\affiliation{Dipartimento di Ingegneria dell'Informazione, 
Universit\`a di Pisa, Via Caruso 16, I-56122 Pisa, Italy.}

\begin{abstract}
The solution of differential problems, and in particular of quantum wave
equations, can in general be performed both in the direct and in the
reciprocal space. 
However, to achieve the same accuracy, direct-space finite-difference 
approaches usually involve handling larger algebraic problems with respect 
to the approaches based on the Fourier transform in reciprocal
space. This is the result of the errors that direct-space discretization
formulas introduce into the treatment of derivatives.
Here, we propose an approach, relying on a set of sinc-based functions,
that allows us to achieve an exact representation of the derivatives in the 
direct space and that is equivalent to the solution in the reciprocal 
space.
We apply this method to the numerical solution of the Dirac equation in an
armchair graphene nanoribbon with a potential varying only in the transverse
direction.
\end{abstract}

\maketitle

\section{Introduction}
\label{introduction}

With the progressive downscaling of electron devices, whose size has reached
well into the nanoscale, the quantum simulation of charge transport has become
increasingly important. This study requires the numerical solution of
the quantum wave equation for the
charge carriers in the device.

In the case of periodic boundary conditions, the solution in the Fourier 
domain represents a spontaneous alternative to a study in the direct space.
In a Fourier analysis, all the functions appearing in the equation
are replaced by their Fourier expansions, and the Fourier coefficients of
the wave function become the unknowns of the problem, 
which in direct space approaches are instead the values of the wave function 
at the points of a discretization grid. 

The main advantage of this technique is the correct 
treatment of the derivatives. In a reciprocal space analysis, each of 
the complex exponential functions appearing in the Fourier expansions can
be derived exactly, without any approximation. On the contrary, in a standard
finite difference solution, the
derivatives are replaced with finite difference approximations involving a
certain number of samples of the function. This approximation introduces a
distortion in the dispersion relation, which severely limits the accuracy of
high-order solutions. The discretization error decreases if a finer grid is
used, at the expense of an increase of the computational effort. 
Moreover, in some cases,
such as the solution of the Dirac equation for massless fermions,
the discretization of the equation on a direct space grid gives rise to
problems such as the appearance of spurious solutions, or fermion doubling, 
unless proper, non-standard discretization techniques are
applied,~\cite{stacey1982,bender1983,susskind1977} while these issues
do not appear using a Fourier solution technique. The solution in the
reciprocal space is especially convenient when a continuum, envelope function
description of the device is adopted and a slowly varying potential
(containing only a limited number of Fourier components) is considered.

Here, we illustrate how the use of a particular set of periodic basis
functions, obtained from the periodic repetition of sinc (cardinal sine)
functions, makes it possible to obtain an exact description of the derivatives 
in the direct space and a solution approach equivalent to that in the 
reciprocal space. This approach allows solving the differential problem
in a very efficient way in the
direct space, without the need to switch to the reciprocal space and
vice versa.

After a general presentation (in Sec.~\ref{method}) of the method that we
propose, in Secs.~\ref{functionder} and \ref{productsec} we describe how
each term of the differential equation is treated, and in Sec.~\ref{fourier},
we demonstrate the equivalence of the sinc-based method to a Fourier
approach.

Finally, in Sec.~\ref{dirac}, we apply this method to the solution of the
the wave equation in an armchair graphene nanoribbon (described using an
envelope function approach) with a potential varying only in the transverse
direction.
Graphene represents an interesting material that, since its isolation in 
2004~\cite{geim}, has been the focus of a vast theoretical and experimental
research activity~\cite{rise,status,castroneto,experimental,mikhailov,
connolly,iwceconn,acsnano,iwceroche}. Among the many applications that have
been proposed for this material, the possibility to use graphene
nanoribbons for the implementation of nanoelectronic devices~\cite{raza,
schwierz,persp} is particularly intriguing. 
The effective wave equation for the envelope functions, which in common
semiconductors corresponds to the effective mass Schr\"odinger
equation~\cite{yu,mitin} (see, for example,
Refs.~\cite{epl,prl,prb,fnl,cavita}),
in monolayer graphene is represented by the Dirac equation~\cite{ando,kp}.
Its numerical solution in graphene ribbons with an arbitrary potential 
landscape has been studied in a few recent papers~\cite{tworzydlo2008,
hernandez2012,snyman2008}. The solution of the Dirac equation
within a slice with a potential varying only in the transverse direction
can be exploited to perform a transport analysis for an armchair 
nanoribbon with a generic potential, if we subdivide the ribbon into a series
of slices in each of which the potential is approximately constant in the 
longitudinal direction, and then we apply a recursive technique to obtain the 
transmission through the overall device~\cite{icnf2013}.

\section{Mathematical method}
\label{method}

For the sampling theorem, if a function $z(x)$ is band limited with band
$B\le 1/(2\,\Delta)$ (Nyquist criterion), it can be perfectly
reconstructed~\cite{haykin} from its samples taken with a sampling
interval $\Delta$
\begin{equation}
z(x)=\sum_{n=-\infty}^{+\infty}z(n\,\Delta)\,
\sinc\left(\frac{x-n\Delta}{\Delta}\right)\,,
\end{equation}
where the sinc function is defined as $\sinc(x)\equiv\sin(\pi x)/(\pi x)$.

\begin{figure}
\begin{center}
\includegraphics[width=8cm]{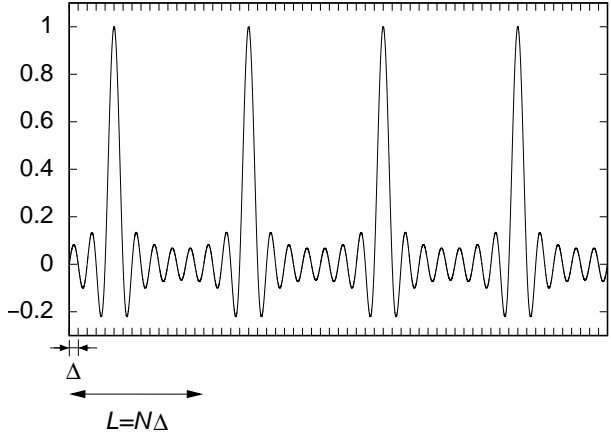}
\caption{Four periods of the function $g_{\ell,\Delta}(x)$
for $N=15$ and $\ell=5$.}
\label{g}
\end{center}
\end{figure}

Moreover, if $z(x)$ is periodic with period $L$ and we take $N$ samples within
a period (and thus $L=N\,\Delta$), we have that
\begin{eqnarray}
z(x)&=&\sum_{\ell=0}^{N-1}\sum_{\eta=-\infty}^{+\infty}
z((\ell+\eta N)\Delta)\,\,
\sinc\left(\frac{x-(\ell+\eta N)\Delta}{\Delta}\right)\nonumber\\
&=&\sum_{\ell=0}^{N-1}z(\ell\,\Delta)\sum_{\eta=-\infty}^{+\infty}
\sinc\left(\frac{x-(\ell+\eta N)\Delta}{\Delta}\right)\nonumber\\
&=&\sum_{\ell=0}^{N-1}z(\ell\,\Delta)g_{\ell,\Delta}(x)\,,
\end{eqnarray}
where we have exploited the periodicity of $z(x)$ and we have defined the
function (periodic with period $L=N\,\Delta$)
\begin{equation}
g_{\ell,\Delta}(x)\equiv \sum_{\eta=-\infty}^{+\infty}
\sinc\left(\frac{x-(\ell+\eta N)\Delta}{\Delta}\right)
\end{equation}
(in Fig.~\ref{g}, we represent 4 periods of the function $g_{\ell,\Delta}(x)$
for $N=15$ and $\ell=5$).

If we define the scalar product between two functions $\psi_1(x)$ and
$\psi_2(x)$ as 
\begin{equation}
\langle \psi_1(x) | \psi_2(x) \rangle=
\frac{1}{\Delta}\int_0^L\psi_1^*(x)\psi_2(x) d\,x \,,
\label{product}
\end{equation}
we can prove (see Appendix~\ref{orthonormality}) that the set of functions
$g_{\ell,\Delta}(x)$ is orthonormal,
i.e. $\langle g_{j,\Delta}(x) | g_{\ell,\Delta}(x) \rangle=\delta_{j,\ell}$
(where $\delta$ is the Kronecker delta).

Here, we consider a wave equation defined on a domain $[0,L]$
and with periodic boundary conditions. The solutions of this differential
equation (i.e., the wave functions), extended by periodicity with period
$L$, $\varphi(x)$ (with $\varphi: \mathbb{R} \to \mathbb{C}$)
are elements of the infinite-dimensional Hilbert space (with scalar product
(\ref{product})) $L^2[0,L]$ of the $L$-periodic functions with finite
norm and are continuous up to the $(\tau-1)$th derivative (if the potential
is bounded, $\tau$ being the order of the differential equation).
In order to solve the problem numerically, here we reformulate it
onto the finite-size subspace generated by the $N$ basis functions $g$ that
we have just defined. This allows to preserve the smoothness of the
wave functions and, in particular, to numerically approximate the $N$
slowest-varying solutions, the ones we are mainly interested in.
In detail, we approximate each of the terms appearing in the wave equation
as a linear combination of the functions $g_{\ell,\Delta}(x)$, with coefficients
given (in their turn) by linear combinations of the samples of the (unknown)
wave function $\varphi(x)$ at the points $n\,\Delta$ (with $n=0,\ldots,N-1$).
In order to easily arrive at closed-form analytical expressions, 
in our calculations, we assume $N$ to be an odd number, i.e., $N=2\,D+1$
(with $D$ an integer). Then, we project the wave equation onto the functions
$g_{j,\Delta}(x)$, with $j=0,\ldots,N-1$. In this way, exploiting the
orthonormality of the $g$ functions, we recast the wave equation into 
a system of $N$ linear equations in the $N$ unknowns $\varphi(n\,\Delta)$
with $n=0,\ldots,N-1$ (corresponding to the $N \times N$ matrix
representation~\cite{datta} of the wave equation on the chosen set of
basis functions). In Sec.~\ref{fourier}, we will prove that this method
is equivalent to a Fourier-Galerkin analysis with cut-off
at the spatial frequency $D/L$ (and it is thus characterized by the same good
convergence properties that have been proved for the Fourier-Galerkin approach,
if the involved functions are sufficiently well-behaved: see, for
example, Ref.~\cite{canuto}).

In our discussion, we will consider the following terms
of the wave equation: a term proportional
to the unknown wave function $\varphi(x)$, a term proportional to a derivative
of the wave function, and a term proportional to the product between the wave
function $\varphi(x)$ and
a function $f(x)$ (with $f: \mathbb{R} \to \mathbb{C}$, in general)
which in $[0,L]$ represents (or is a function of) the potential
energy and is then extended by periodicity with period $L$.
Other possible terms
appearing in the wave equation will undergo a similar treatment.

\section{Treatment of the wave function and of its derivatives}
\label{functionder}

We assume the unknown periodic wave function $\varphi(x)$ to be a
band limited function with a maximum band $B=D/L$, such that, sampling it
with a step $\Delta$, the Nyquist criterion $B \le 1/(2\,\Delta)$ is satisfied.

Therefore, the term proportional to the unknown wave function is easily
rewritten in the desired form by applying the sampling theorem with a
step $\Delta$
\begin{equation}
\varphi(x)=\sum_{\mu=0}^{N-1}\varphi(\mu\,\Delta)g_{\mu,\Delta}(x)\,.
\label{exp-rho}
\end{equation}

Concerning the term proportional to the derivative of the wave function 
$\varphi(x)$, if $\varphi(x)$ is band limited with a band equal to $B=D/L$,
also
its derivatives have the same property, and thus we can apply the sampling
theorem to them, too, obtaining (for the generic $s$-th derivative):
\begin{equation}
\frac{d^s\,\varphi(x)}{d\,x^s}=\sum_{\mu=0}^{N-1}
\left[\frac{d^s\,\varphi(x)}{d\,x^s}\right]_{x=\mu\,\Delta}g_{\mu,\Delta}(x)\,.
\end{equation}
We now have to express the values of the derivatives in the points
$\mu\,\Delta$ as a function of the samples of $\varphi(x)$.
This can be easily achieved by deriving Eq.~(\ref{exp-rho}). We have that
\begin{eqnarray}
\displaystyle \left[\frac{d^s\,\varphi(x)}{d\,x^s}\right]_{x=\mu\,\Delta}&=&
\left[\frac{d^s}{d\,x^s}
\left(\sum_{\ell=0}^{N-1}
\varphi(\ell\,\Delta)g_{\ell,\Delta}(x)\right)\right]_{x=\mu\,\Delta}\nonumber\\
&=&\displaystyle\sum_{\ell=0}^{N-1}\varphi(\ell\,\Delta)
\sum_{\eta=-\infty}^{+\infty}
\left[\frac{d^s}{d\,x^s}
\sinc\left(\frac{x-(\ell+\eta N)\Delta}
{\Delta}\right)\right]_{x=\mu\,\Delta}\nonumber\\
&=&\sum_{\ell=0}^{N-1}\varphi(\ell\,\Delta)
\alpha_{\mu \ell}^s \,,
\label{alpha}
\end{eqnarray}
where we have defined as $\alpha_{\mu \ell}^s$ the coefficients with which we
have to combine the samples of the unknown wave function $\varphi(x)$ in order
to determine the samples of its $s$th derivative.

In particular, let us develop the calculation for $s=1$ (first derivative) and
for $s=2$ (second derivative).

For $s=1$, we have that
\begin{equation}
\frac{d}{d\,x}
\sinc\left(\frac{x-(\ell+\eta N)\Delta}{\Delta}\right)
=\frac{\cos\left(\pi\,\frac{\displaystyle x-(\ell+\eta N)\Delta}
{\displaystyle \Delta}\right)}
{x-(\ell+\eta N)\Delta}-
\frac{\sin\left(\pi\,\frac{\displaystyle x-(\ell+\eta N)\Delta}
{\displaystyle \Delta}\right)}
{\frac{\displaystyle \pi}{\displaystyle \Delta}
\left(x-(\ell+\eta N)\Delta\right)^2}
\end{equation}
and thus
\begin{equation}
\left[\frac{d}{d\,x}
\sinc\left(\frac{x-(\ell+\eta N)\Delta}
{\Delta}\right)\right]_{x=\mu\,\Delta}=\begin{cases}
\displaystyle 0 & \text{if $\mu-\ell-\eta N=0$}\\[10pt]
\displaystyle \frac{(-1)^{\mu-\ell-\eta N}}{\Delta (\mu-\ell-\eta N)}
& \text{if $\mu-\ell-\eta N \ne 0$.}
\end{cases}
\end{equation}
If we develop the calculation assuming $N$ odd, we have that
\begin{equation}
\alpha_{\mu \ell}^1=\sum_{\eta=-\infty}^{+\infty}\left[\frac{d}{d\,x}
\sinc\left(\frac{x-(\ell+\eta N)\Delta}
{\Delta}\right)\right]_{x=\mu\,\Delta}=\begin{cases}
\displaystyle 0 & \text{if $\mu=\ell$}\\[10pt]
\displaystyle \frac{(-1)^{\mu-\ell}\,\pi}
{N\,\Delta \sin\left(\pi \,\frac{\displaystyle \mu-\ell}
{\displaystyle N}\right)}
& \text{if $\mu \ne \ell$.}
\end{cases}
\label{alpha1}
\end{equation}

Analogously, for $s=2$, we have that
\begin{eqnarray}
&\displaystyle \frac{d^2}{d\,x^2}
\sinc\left(\frac{x-(\ell+\eta N)\Delta}{\Delta}\right)
=-\frac{\pi}{\Delta}
\frac{\sin\left(\pi\,\frac{\displaystyle x-(\ell+\eta N)\Delta}
{\displaystyle \Delta}\right)}
{x-(\ell+\eta N)\Delta}&\nonumber\\
&\displaystyle 
-2\frac{\cos\left(\pi\,\frac{\displaystyle x-(\ell+\eta N)\Delta}
{\displaystyle \Delta}\right)}
{\left(x-(\ell+\eta N)\Delta\right)^2}
+\frac{2\Delta}{\pi}
\frac{\sin\left(\pi\,\frac{\displaystyle x-(\ell+\eta N)\Delta}
{\displaystyle \Delta}\right)}
{\left(x-(\ell+\eta N)\Delta\right)^3}&\nonumber\\
\end{eqnarray}
and thus
\begin{equation}
\left[\frac{d^2}{d\,x^2}
\sinc\left(\frac{x-(\ell+\eta N)\Delta}
{\Delta}\right)\right]_{x=\mu\,\Delta}=\begin{cases}
\displaystyle -\frac{\pi^2}{3\Delta^2} & \text{if $\mu-\ell-\eta N=0$}\\[10pt]
\displaystyle -2\frac{\displaystyle (-1)^{\mu-\ell-\eta N}}
{\displaystyle \Delta^2 (\mu-\ell-\eta N)^2}
& \text{if $\mu-\ell-\eta N \ne 0$.}
\end{cases}
\end{equation}
Assuming $N$ odd, we obtain that
\begin{eqnarray}
\alpha_{\mu \ell}^2
&=&\displaystyle \sum_{\eta=-\infty}^{+\infty}\left[\frac{d^2}{d\,x^2}
\sinc\left(\frac{x-(\ell+\eta N)\Delta}
{\Delta}\right)\right]_{x=\mu\,\Delta}\nonumber\\[10pt]
&=&\displaystyle \begin{cases}
\displaystyle \frac{\pi^2}{3N^2\Delta^2}(1-N^2) & \text{if $\mu=\ell$}\\[10pt]
\displaystyle -\frac{2\pi^2}{N^2\Delta^2}(-1)^{\mu-\ell}
\frac{\cos\left(\frac{\displaystyle (\mu-\ell)\pi}{\displaystyle N}\right)}
{\sin^2\left(\frac{\displaystyle (\mu-\ell)\pi}{\displaystyle N}\right)}
& \text{if $\mu \ne \ell$.}
\end{cases}
\label{alpha2}
\end{eqnarray}

Notably, these expressions coincide with those of the so-called SLAC
derivative technique~\cite{drell}, which were
obtained~\cite{foerster}
switching to the reciprocal space, operating in that domain and then
transforming back (analogously to what we do in Sec.~\ref{fourier},
where we compare our method with the Fourier one).

\section{Treatment of the product term}
\label{productsec}

The last term that we consider in our analysis is the one proportional to the
product between the function $f(x)$ (the potential energy or a function of it)
and the wave function $\varphi(x)$.

To a first approximation (simplified sinc-based approach), we can proceed as
if the product term were band limited with band $B=D/L$ and thus directly
express it as a linear combination of the functions $g_{\ell,\Delta}(x)$, with
coefficients given by the samples of the product term at the points $n\,\Delta$
(with $n=0,\ldots,N-1$). Operating in this way, in the final system of
equations this term gives only a diagonal contribution consisting of its
samples at the points $n\,\Delta$ (analogously to Ref.~\cite{foerster}).
However, since we have assumed a band $B$ for $\varphi(x)$, the same
assumption is in general not verified for the product term (unless a constant
potential function is considered); and for high-order solutions,
this introduces a discrepancy with respect to the solutions from a 
Fourier analysis.

In the following, we discuss a better approximation (advanced sinc-based
approach) that makes this direct-space method equivalent to the
Fourier one.

We introduce an odd (for analytical convenience) positive integer
number $M$, which accounts for the fact that in general the exploitation of
more samples of the potential than just those at the positions where we want
to evaluate the wave function $\varphi$ can be useful.
In the calculation we include the samples of the potential 
function $f(x)$ at the $N\,M$ ($=2\,Q+1$, with $Q$ a positive integer)
points taken at intervals multiple of $\Delta'=L/(N\,M)=\Delta/M$.
We replace the function $f(x)$ with the function
\begin{eqnarray}
\tilde f(x)&=&
\sum_{m=0}^{MN-1}f\left(m\,\Delta'\right)
\sum_{\eta=-\infty}^{+\infty}
\sinc\left(\frac{\displaystyle x-(m+\eta MN)\Delta'}
{\displaystyle \Delta'}\right)\nonumber\\
&=&\sum_{m=0}^{MN-1}f\left(m\,\Delta'\right)
g_{m,\Delta'}(x)
\label{tildef}
\end{eqnarray}
(with a maximum band $B'=Q/L$), obtained by reconstruction from its $N\,M$
samples taken at intervals $\Delta'$.

Assuming for the wave function the expression (\ref{exp-rho}), we can
therefore write the product term as:
\begin{equation}
h(x)=\tilde f(x)\varphi(x)=
\left[\sum_{m=0}^{MN-1}f\left(m\,\Delta'\right) g_{m,\Delta'}(x)\right]
\left[\sum_{\ell=0}^{N-1}\varphi(\ell\,\Delta)g_{\ell,\Delta}(x)\right]\,.
\label{h}
\end{equation}

In order to obtain the final matrix representation of our wave equation,
we have to compute only the projections of $h(x)$ on the functions
$g_{\mu,\Delta}(x)$ (with $\mu=0,\ldots,N-1$), which corresponds to
approximating $h(x)$ with
\begin{equation}
h_0(x)=\sum_{\mu=0}^{N-1}\langle g_{\mu,\Delta}(x)|h(x)\rangle\, g_{\mu,\Delta}(x)
=\sum_{\mu=0}^{N-1}\left[\frac{1}{\Delta}\int_{0}^{L}g_{\mu,\Delta}^*(\chi)
h(\chi) d\,\chi\right]g_{\mu,\Delta}(x)\,.
\label{h0}
\end{equation}
These projections are equal (exploiting Eq.~\eqref{h}) to
\begin{eqnarray}
&& \displaystyle \frac{1}{\Delta}\int_{0}^{L}
g_{\mu,\Delta}^*(\chi)h(\chi) d\,\chi\nonumber\\
&& \displaystyle
=\sum_{\ell=0}^{N-1}\Bigg\{ \varphi(\ell\,\Delta)
\sum_{m=0}^{MN-1} \bigg[ f\left(m\,\Delta'\right)
\frac{1}{\Delta}\int_{0}^{L}
g_{\mu,\Delta}^*(\chi)g_{m,\Delta'}(\chi)g_{\ell,\Delta}(\chi)d\,\chi \bigg]\Bigg\}
\nonumber\\
&& \displaystyle 
=\sum_{\ell=0}^{N-1}\Bigg\{ \varphi(\ell\,\Delta)
\frac{1}{M N^2}\sum_{m=0}^{MN-1}f\left(m\,\Delta'\right)
\nu_{\mu \ell m} \Bigg\}\nonumber\\
&& \displaystyle 
=\sum_{\ell=0}^{N-1}\varphi(\ell\,\Delta)\beta_{\mu \ell}\,,
\label{pro}
\end{eqnarray}
where the coefficients $\beta_{\mu \ell}$ are given by
\begin{equation}
\beta_{\mu \ell} \equiv
\frac{1}{M N^2}\sum_{m=0}^{MN-1}f\left(m\,\Delta'\right)\nu_{\mu \ell m}\,.
\label{beta}
\end{equation}
With some analytical calculations  (a possible procedure will be briefly
described in the second part of Appendix~\ref{productnotes}) it is possible
to find a closed form for
\begin{equation}
\nu_{\mu \ell m} \equiv M N^2\frac{1}{\Delta}\int_{0}^{L}
g_{\mu,\Delta}^*(\chi)g_{m,\Delta'}(\chi)g_{\ell,\Delta}(\chi)d\,\chi\,.
\label{nu}
\end{equation}
In detail, if $M\ne 1$ we have that
\begin{equation}
\nu_{\mu \ell m}=\lambda_1(m,\mu)\,\lambda_1(m,\ell)
\label{mne1}
\end{equation}
with
\begin{equation}
\lambda_1(m,\mu)=\begin{cases}
\displaystyle N & \text{if $\mu M=m$}\\[10pt]
\displaystyle
\frac{\displaystyle\sin\left(\pi\left(\mu-\frac{m}{M}\right)\right)}
{\displaystyle\sin\left(\frac{\pi}{N}\left(\mu-\frac{m}{M}\right)\right)}
 & \text{if $\mu M \ne m$.}
\end{cases}
\end{equation}
Instead, if $M=1$ (and thus $\Delta'=\Delta$) we have that
\begin{equation}
\nu_{\mu \ell m}=\begin{cases}
\displaystyle \lambda_2(m,\mu) & \text{if $\ell =m$}\\[10pt]
\displaystyle
\frac{\displaystyle (-1)^{\ell-m}\,
\left(\lambda_3(\ell,\mu)-\lambda_3(m,\mu)\right)}
{\displaystyle
\sin\left(\frac{\pi}{N}\,(\ell-m)\right)}& \text{if $\ell \ne m$}
\end{cases}
\label{meq1}
\end{equation}
with
\begin{equation}
\lambda_2(m,\mu)=\begin{cases}
\displaystyle 3D^2+3D+1 & \text{if $m=\mu$}\\[10pt]
\displaystyle
\frac{\displaystyle \sin^2\left(\frac{\pi}{N}\,D\,(\mu-m)\right)}
{\displaystyle \sin^2\left(\frac{\pi}{N}\,(\mu-m)\right)}
 & \text{if $m \ne \mu$}
\end{cases}
\end{equation}
and
\begin{equation}
\lambda_3(m,\mu)=\begin{cases}
\displaystyle 0 & \text{if $m=\mu$}\\[10pt]
\displaystyle
\frac{\displaystyle \cos\left(\frac{\pi}{N}\,(\mu-m)\right)-(-1)^{\mu-m}}
{\displaystyle 2\sin\left(\frac{\pi}{N}\,(\mu-m)\right)}
 & \text{if $m \ne \mu$.}
\end{cases}
\end{equation}

\section{Correspondence with the Fourier analysis}
\label{fourier}

The differential problem we are
interested in can alternatively be solved using a classical solution method in
the transformed domain. We can replace the wave function
$\varphi(x)$ and the potential energy function $f(x)$ with their truncated
Fourier expansions. In detail, we can consider only the lowest (in modulus)
$N$ spatial frequencies of $\varphi(x)$ (i.e., the spatial frequencies
$\ell/L$ with $|\ell| \le D$, where $N=2\,D+1$) and we can numerically compute
the lowest-order $M\,N$ Fourier coefficients of $f(x)$ by means of a discrete
Fourier transform (DFT) of its $M\,N$ samples within the period $L$.
Then, we can project the resulting equation onto the $N$ basis functions
$\exp(i2\pi j x/L)$ with $|j| \le D$, thereby
recasting the problem into an $N\times N$ system of linear equations where
the Fourier coefficients of the wave function are the unknowns.
In this way, we approximate the infinite-dimensional problem to a finite-size
matrix problem, disregarding the Fourier components corresponding to
frequencies greater in modulus than $D/L$, both for the unknown wave function 
and for all the terms appearing in the equation.

We will now show that the advanced sinc-based approach 
described in the previous sections is equivalent to this
Fourier-based solution technique.

As we have seen,
a periodic function $\tilde z(x)$ with maximum band $D/L$ can be expressed
in terms of its $N=2\,D+1$ samples at the points $n\,\Delta$ within the
period $L$ as follows:
\begin{equation}
\tilde z(x)=\sum_{\ell=0}^{N-1}\tilde z(\ell\,\Delta)g_{\ell,\Delta}(x)\,.
\end{equation}
Therefore, we can compute its Fourier series coefficients $\tilde Z_p$
noting that the Fourier coefficients $[G_{\ell,\Delta}]_p$ of
the function $g_{\ell,\Delta}(x)$ are given by (exploiting the
definition of $g_{\ell,\Delta}(x)$ as a periodic repetition of sinc
functions~\cite{haykin})
\begin{equation}
\left[G_{\ell,\Delta}\right]_p=\frac{1}{L}\mathcal{F}
\left[\sinc\left(
\frac{x-\ell\Delta}{\Delta}\right)
\right]_{f=\frac{\scriptstyle p}{\scriptstyle L}}
=\frac{1}{N}e^{-i2\pi p \frac{\scriptstyle \ell}{\scriptstyle N}}
\label{trasfg}
\end{equation}
for $|p| \le D$, and 0 otherwise (with $\mathcal{F}$ the Fourier
transform) and thus
\begin{equation}
\tilde Z_p=
\sum_{\ell=0}^{N-1}\tilde z(\ell\,\Delta)\left[G_{\ell,\Delta}\right]_p= 
\frac{1}{N}\sum_{\ell=0}^{N-1}\tilde z(\ell\,\Delta)
e^{-i2\pi p \frac{\scriptstyle \ell}{\scriptstyle N}}
\label{coeff}
\end{equation}
for $|p| \le D$, and 0 otherwise.

It is apparent that Eq.~\eqref{coeff} corresponds to the DFT of $\tilde z(x)$.

Therefore the DFT of a function $z(x)$ computed from its $N=2\,D+1$ samples
coincides with the exact Fourier series of the function $\tilde z(x)$ with
maximum band $D/L$ that has the same samples as $z(x)$ in the period $L$.

In particular, this consideration can be applied
to the potential function $f(x)$: calculating its DFT on the $M\,N$
samples (as we do in the Fourier analysis)
corresponds exactly to substituting $f(x)$ with the function
$\tilde f(x)$ of Eq.~(\ref{tildef}) (as we do in our advanced sinc-based
approach).

Moreover, as we have seen, when operating in the frequency domain,
we disregard the frequency components outside the interval $[-D/L,D/L]$
for all the terms of the differential problem. 
Let us discuss, for each term, the equivalent of this frequency cut-off
in the direct domain.

Limiting the Fourier components of the unknown wave function
$\varphi(x)$ to frequencies that have a modulus less than $D/L$ corresponds
exactly to the assumption, we have made while operating in the direct space,
that $\varphi(x)$ is band limited with band $D/L$ (which has allowed us to
use the sampling theorem with $N=2\,D+1$ samples).

The same consideration is valid for the derivatives of $\varphi(x)$.
The expressions that we have found for the derivatives can actually
be obtained also from a reciprocal space analysis. Indeed, if 
$\varphi(x)$ is periodic and we assume it to be band limited with band
$B=D/L$ with Fourier coefficients $a_p$, we can write it and its
derivatives in the form
\begin{equation}
\varphi(x)=\sum_{p=-D}^{D} a_p 
e^{i2\pi p \frac{\scriptstyle x}{\scriptstyle L}}\,,
\end{equation}
\begin{equation}
\frac{d^s \varphi(x)}{d x^s}=
\sum_{p=-D}^{D} \left(i 2 \pi \frac{p}{L}\right)^s
a_p e^{i2\pi p \frac{\scriptstyle x}{\scriptstyle L}}\,.
\label{derfreq}
\end{equation}
The function $\exp(i2\pi p x/L)$ with $|p|\le D$
can be seen as a band limited function with band $B=D/L$ and thus
expressed in terms of its $N$ samples in the period $L$:
\begin{equation}
e^{i2\pi p \frac{\scriptstyle x}{\scriptstyle L}}=
\sum_{\mu=0}^{N-1}e^{i2\pi p \frac{\scriptstyle \mu}{\scriptstyle N}}
g_{\mu,\Delta}(x)\,.
\end{equation}
Substituting this expression into Eq.~(\ref{derfreq}) we obtain
\begin{equation}
\frac{d^s \varphi(x)}{d x^s}=
\sum_{\mu=0}^{N-1}\left[\sum_{p=-D}^{D} \left(i 2 \pi \frac{p}{L}\right)^s
a_p e^{i2\pi p \frac{\scriptstyle \mu}{\scriptstyle N}}\right]g_{\mu,\Delta}(x)\,.
\end{equation}
The value of the derivative for $x=\mu\,\Delta$ is the quantity between
square brackets.
Therefore (if we express $a_p$ using Eq.~(\ref{coeff})), we have that
\begin{eqnarray}
&&\left[\frac{d^s \varphi(x)}{d x^s}\right]_{x=\mu\,\Delta}\nonumber\\
&&=\sum_{p=-D}^{D} \left(i 2 \pi \frac{p}{L}\right)^s
\left(\frac{1}{N}\sum_{\ell=0}^{N-1}\varphi(\ell\,\Delta)
e^{-i2\pi p \frac{\scriptstyle \ell}{\scriptstyle N}}\right)
e^{i2\pi p \frac{\scriptstyle \mu}{\scriptstyle N}}\nonumber\\
&&=\sum_{\ell=0}^{N-1}\varphi(\ell\,\Delta)
\left[\left(\frac{i 2 \pi}{(2D+1)\Delta}\right)^s
\frac{1}{2D+1}
\sum_{p=-D}^{D}p^s
e^{i2\pi p \frac{\scriptstyle \mu-\ell}{\scriptstyle 2D+1}}\right]\nonumber\\
&&=\sum_{\ell=0}^{N-1}\varphi(\ell\,\Delta)\alpha_{\mu \ell}^s\,.
\label{alphaf}
\end{eqnarray}
It is easy to verify that the expression for $\alpha_{\mu \ell}^s$ in
Eq.~(\ref{alphaf}) yields exactly the same closed-form results that can be
found operating in the direct space [for example, for $s=1$ and $s=2$,
those reported in Eqs.~(\ref{alpha1}) and (\ref{alpha2})].
Indeed, the
expression in Eq.~(\ref{alphaf}) can be easily obtained also by substituting
the Fourier expansion (\ref{ser1}) of $g_{\ell,\Delta}(x)$ into the definition
$\alpha_{\mu \ell}^s=[d^s\,g_{\ell,\Delta}(x)/d\,x^s]_{x=\mu\,\Delta}$
of Eq.~(\ref{alpha}).

Finally, let us consider the effect of the frequency cut-off
on the product term $h(x)$.
Neglecting the frequency components of this function outside the interval
$[-D/L,D/L]$ corresponds to limiting its band to the range
$[-1/(2\Delta),1/(2\Delta)]$, because $1/(2\Delta)=(2D+1)/(2L)=(D/L)+[1/(2L)]$
but the product term, being periodic with period $L$, contains only frequency
components multiple of $1/L$.
If $H(f)$ is the Fourier transform of the function $h(x)$
($f$ being the spatial frequency), performing this frequency cut-off
is equivalent to considering, in the Fourier domain, a spectrum
$H_0(f)=\rect(\Delta f) H(f)$, where $\rect(\Delta f)$ is a function equal
to 1 between $-1/(2\Delta)$ and $1/(2\Delta)$, and to 0 outside this interval.
The inverse Fourier transform of $H_0(f)$ is equal to
\begin{equation}
h_0(x)=\frac{1}{\Delta}\int_{-\infty}^{+\infty}
\sinc\left(\frac{x-\chi}{\Delta}\right)h(\chi) d\,\chi\,.
\end{equation}
Since the periodic function $h_0(x)$ is band limited with band 
$B\le 1/(2\Delta)$, we can express it in terms of its samples taken with
sampling interval $\Delta$
\begin{eqnarray}
h_0(x)&=&\sum_{\ell=0}^{N-1}\left[
\frac{1}{\Delta}\int_{-\infty}^{+\infty}
\sinc\left(\frac{\chi-\ell\Delta}{\Delta}\right)h(\chi) d\,\chi\right]
g_{\ell,\Delta}(x)\nonumber\\
&=&\sum_{\ell=0}^{N-1}\left[
\frac{1}{\Delta}\int_{0}^{L}
g_{\ell,\Delta}^*(\chi)h(\chi) d\,\chi\right]
g_{\ell,\Delta}(x)
\end{eqnarray}
[where we have exploited the fact that the sinc function is even and
Eq.~(\ref{changedom})].
This exactly corresponds to what we have done in the direct space [see
Eq.~(\ref{h0})]. Indeed, we can verify that the expression of the
product term obtained operating in the reciprocal domain is equivalent to
that achieved in the direct space (see Appendix~\ref{productnotes}).

Therefore, with the approach we have described, the solutions in the direct
and in the reciprocal space are equivalent and correspond to linear systems
of the same size.

In order to further clarify this equivalence,
let us notice that the periodic and band limited function $\varphi(x)$ can be
equivalently expressed, using both the Fourier expansion and the
sampling theorem, in the following two ways:
\begin{equation}
\varphi(x)=\sum_{p=-D}^{D}(\sqrt{N}\,a_p)
\frac{e^{i2\pi p \frac{\scriptstyle x}{\scriptstyle L}}}{\sqrt{N}}=
\sum_{\ell=0}^{2D}\varphi(\ell\,\Delta)g_{\ell,\Delta}(x)\,.
\end{equation}
Two different sets of orthonormal basis functions have been used in this
equation: the functions $\exp(i2\pi p x/L)/\sqrt{N}$ and the
functions $g_{\ell,\Delta}(x)$ [with this choice, the exponentials have been 
properly normalized with respect to the scalar
product (\ref{product})]. Performing the solution in the reciprocal domain
corresponds to using the former basis set, while performing the solution in
the direct space corresponds to using the latter basis set.\\
It is possible to switch from one basis to the other 
noticing that each of the exponential functions that appear in the Fourier
expansion (being band limited with band $B\le 1/(2\Delta)$) can
be expressed in terms of its samples taken with sampling interval $\Delta$
in this way
\begin{equation}
\frac{e^{i2\pi p \frac{\scriptstyle x}{\scriptstyle L}}}{\sqrt{N}}=
\sum_{\ell=0}^{2D}
\frac{e^{i2\pi p \frac{\scriptstyle \ell}{\scriptstyle N}}}{\sqrt{N}}
g_{\ell,\Delta}(x)=
\sum_{\ell=0}^{2D}T_{\ell p}g_{\ell,\Delta}(x)\,.
\end{equation}
The matrix $T$, made up of the $T_{\ell p}$ elements, can be used to switch
from the Fourier basis to the direct-space one.\\ 
On the other hand, using Eq.~\eqref{trasfg}, we can write the Fourier
expansion of $g_{\ell,\Delta}(x)$ as
\begin{equation}
g_{\ell,\Delta}(x)=\sum_{p=-D}^{D}\left[G_{\ell,\Delta}\right]_p
e^{i2\pi p \frac{\scriptstyle x}{\scriptstyle L}}=
\sum_{p=-D}^{D}\sqrt{N} \left[G_{\ell,\Delta}\right]_p
\frac{e^{i2\pi p \frac{\scriptstyle x}{\scriptstyle L}}}{\sqrt{N}}
\end{equation}
and thus the matrix $\Theta=T^{\dagger}$, made up of the elements
$\Theta_{p \ell}=\sqrt{N} [G_{\ell,\Delta}]_p=
\exp(-i2\pi p \ell/N)/\sqrt{N}=T_{\ell p}^*\,$, operates
the change from the direct-space basis to the Fourier one.

In particular, if the matrix of the direct space linear system is $M_d$ and
the matrix of the reciprocal-space linear system is $M_r$, we have that
$M_d=T\,M_r\,T^{-1}=T\,M_r\,T^{\dagger}$.

\section{Application to the solution of the Dirac equation in an armchair
graphene nanoribbon}
\label{dirac}

Transport in graphene is described, within an envelope function approach,
by the Dirac equation~\cite{kp}. When considering graphene nanoribbons, 
at the effective edges of the ribbon (i.e., at the lattice sites just
outside the ribbon), we have to enforce Dirichlet boundary conditions,
which couple the envelope functions corresponding to the two inequivalent
Dirac points $\vec K$ and $\vec K'$. 
In particular, here we take into consideration the solution of the Dirac
equation in the case in which the potential in the ribbon depends only
on the transverse co-ordinate, i.e., in which the potential does not vary in the
longitudinal direction. In this case, the four envelope functions
$F_{S \vec P}(\vec r)$ (corresponding to the two Dirac points
$\vec P=\vec K,\vec K'$ and to the two sublattices $S=A,B$) can be written
as the product of a confined component in the transverse $y$ direction
$\Phi_{S \vec P}(y)$ and of a propagating wave in the longitudinal
$x$ direction (with longitudinal wave vector $\kappa_x$):
$F_{S \vec P}(\vec r)=\Phi_{S \vec P}(y)\,\exp(i \kappa_x x)$.

As shown in Refs.~\cite{pt,iwce2010}, the problem can be reformulated
into a differential problem with periodic boundary conditions on a doubled
domain ($2\tilde W$ instead of $\tilde W$, which represents the effective
width of the ribbon, i.e., the distance between the two effective edges
where the Dirichlet boundary conditions had to be enforced in the original
problem).
This new problem can be expressed in the following form:
\begin{equation}
\begin{cases}
\displaystyle
\sigma_z \left(\frac{d\,\vec\varphi(y)}{d\,y}+i\tilde K \vec\varphi(y)\right)
+\sigma_x f(y)\vec\varphi(y)=-\kappa_x \vec\varphi(y)\\[10pt]
\displaystyle
\vec\varphi(2\tilde W)=\vec\varphi(0)\,,
\end{cases}
\end{equation}
where the $\sigma$'s are the Pauli matrices and $\vec\varphi(y)$ is a
two-component function directly related to the four envelope functions
of graphene in the following way:
\begin{equation}\label{varphi}
\vec\varphi(y)=
\begin{cases}
\displaystyle
e^{-i \tilde K y}
\left[\begin{array}{c}
\Phi_{A \vec K}(y)\\
\Phi_{B \vec K}(y)
\end{array}\right] &
\text{if $y \in [0,\tilde W]$}\\[10pt]
\displaystyle e^{i \tilde K (2\tilde W-y)}\,i\,
\left[\begin{array}{c}
\Phi_{A \vec K'}(2\tilde W-y)\\
\Phi_{B \vec K'}(2\tilde W-y)
\end{array}\right] &
\text{if $y \in [\tilde W,2\tilde W]$.}
\end{cases}
\end{equation}
Moreover, $K=4\pi/(3a)$ (where $a$ is the graphene lattice constant) is the
modulus of the transversal co-ordinate of the Dirac points and
$\tilde K=K-\,{\rm round}\,(K /(\pi/\tilde W))\cdot\pi/\tilde W$ is a quantity
such that $\exp(i 2 \tilde K \tilde W)=\exp(i 2 K \tilde W)$ and that
$|\tilde K|<\pi/\tilde W$. We consider $\tilde K$ instead of $K$, in such
a way that slowly varying $\vec\varphi(y)$ functions correspond to
slowly varying transverse components of the envelope functions, i.e., those 
in which we are most interested. Additionally,
$f(y)=(U(\tilde W-|\tilde W-y|)-E)/\gamma$ is the
potential term, where $U(y)$ is the potential energy, $E$ is the Fermi energy,
$\gamma=\hbar v_F$, $\hbar$ is the reduced Planck constant, and $v_F$ is the
Fermi velocity in  graphene. In this formulation, the problem has to be solved
over the domain $[0,2\,\tilde W]$ and the boundary condition for 
$\vec\varphi(y)$ is periodic.

The periodic boundary condition makes it possible to solve the problem in the
Fourier domain or equivalently in the basis of $g$ functions.
Working in the basis of $g$ functions, we have rewritten the equation
as follows:
\begin{eqnarray}
&&\displaystyle \sigma_z \left[\sum_{\mu=0}^{2D}
\left(\sum_{\ell=0}^{2D}\vec\varphi(\ell\,\Delta) \alpha_{\mu \ell}^1\right)
g_{\mu,\Delta}(x)
+i\tilde K \sum_{\mu=0}^{2D}\vec\varphi(\mu\,\Delta)g_{\mu,\Delta}(x)
\right]\nonumber\\
&&\displaystyle +\sigma_x  
\sum_{\mu=0}^{2D}\left(\sum_{\ell=0}^{2D}\vec\varphi(\ell\,\Delta)
\beta_{\mu \ell}\right) g_{\mu,\Delta}(x)=-\kappa_x 
\sum_{\mu=0}^{2D}\vec\varphi(\mu\,\Delta)g_{\mu,\Delta}(x).\nonumber\\
\end{eqnarray}
Projecting this equation (using the scalar product (\ref{product}))
onto the generic function $g_{j,\Delta}(x)$, we obtain
\begin{equation}
\sigma_z \left[
\left(\sum_{\ell=0}^{2D}\vec\varphi(\ell\,\Delta) \alpha_{j \ell}^1\right)
+i\tilde K \vec\varphi(j\,\Delta)\right]+\sigma_x  
\left(\sum_{\ell=0}^{2D}\vec\varphi(\ell\,\Delta)\beta_{j \ell}\right)
=-\kappa_x 
\vec\varphi(j\,\Delta)
\end{equation}
for $j=0,\ldots,2D$, or equivalently
\begin{equation}
\sum_{\ell=0}^{2D}
\left[\sigma_z (\alpha_{j \ell}^1+i\tilde K \delta_{j,\ell})+\sigma_x  
\beta_{j \ell}\right]\vec\varphi(\ell\,\Delta)
=-\kappa_x \vec\varphi(j\,\Delta)
\label{system}
\end{equation}
for $j=0,\ldots,2D$.
These $N=2\,D+1$ equations can be written in matrix form as an eigenproblem
with eigenvalues $-\kappa_x$ and eigenvectors containing the values
$\vec\varphi(j\,\Delta)$ of the unknown function $\vec\varphi(y)$ at the
$N$ points of the considered grid.

In order to adopt the first approximate approach for the product term
described at the beginning of Sec.~\ref{productsec}, we
just have to replace $\beta_{j \ell}$ with $f(j\,\Delta)\delta_{j,\ell}$
in Eq.~(\ref{system}).

\begin{figure}
\begin{center}
\includegraphics[width=8cm]{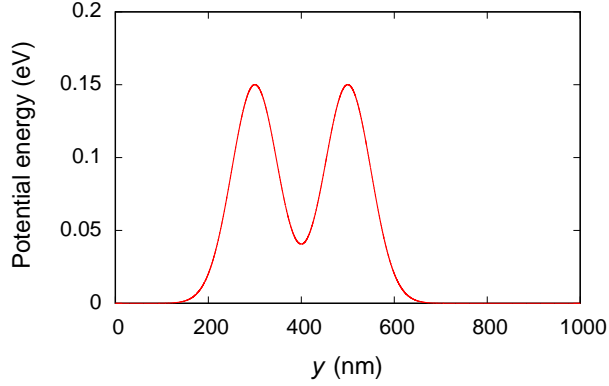}
\caption{Potential energy considered in the armchair ribbon.}
\label{potential}
\end{center}
\end{figure}

\begin{figure}
\begin{center}
\includegraphics[width=8cm]{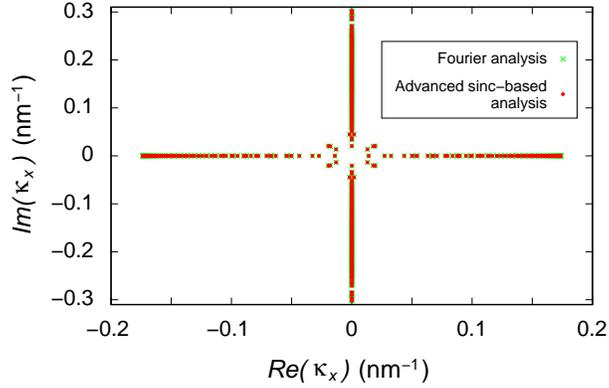}
\caption{Position on the Gauss plane of all the longitudinal wave
vectors $\kappa_x$ obtained with the Fourier analysis and with our advanced
sinc-based approach.}
\label{eig}
\end{center}
\end{figure}

\begin{figure}
\begin{center}
\includegraphics[width=8cm]{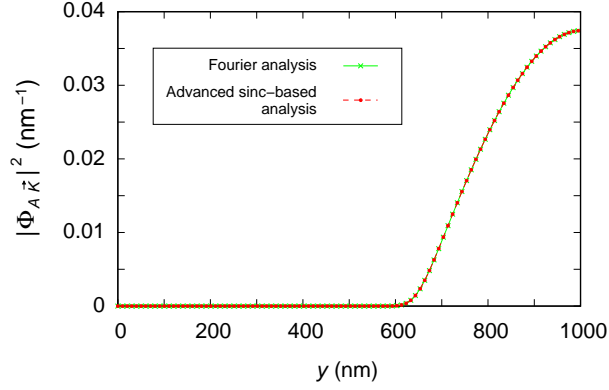}
\caption{Envelope function $\Phi_{A \vec K} (y)$
corresponding to the eigenvalue with the largest real part, computed with
a Fourier analysis and with our advanced sinc-based approach.}
\label{phia}
\end{center}
\end{figure}

\begin{figure}
\begin{center}
\includegraphics[width=8cm]{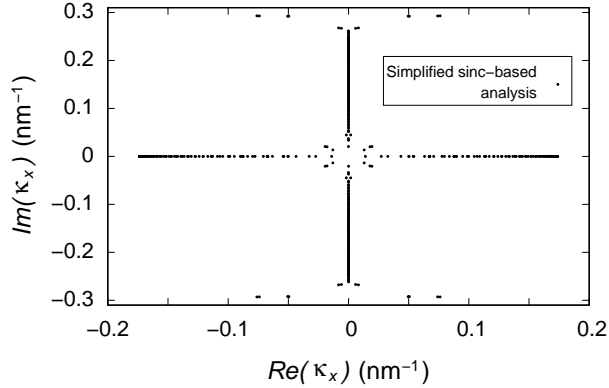}
\caption{Position on the Gauss plane of all the longitudinal wave
vectors $\kappa_x$ obtained with the simplified sinc-based approach.
The highest order wave vectors strongly differ from
those obtained with a Fourier analysis (see Fig.~\ref{eig}).}
\label{eig1}
\end{center}
\end{figure}

For example, we have considered a potential $U(y)$ (represented in
Fig.~\ref{potential}) equal to the sum of two Gaussian functions:
\begin{equation}
U(y)=0.15~{\rm eV}\cdot
e^{-\frac{\scriptstyle (y-300\>{\rm nm})^2}{\scriptstyle 5000\>{\rm nm}^2}}+
0.15~{\rm eV}\cdot
e^{-\frac{\scriptstyle (y-500\>{\rm nm})^2}{\scriptstyle 5000\>{\rm nm}^2}}
\end{equation}
in a 8131 dimer line ($\sim$1~$\mu$m) wide armchair nanoribbon,
considering an electron injection energy $E=0.1$~eV.
In Figs.~\ref{eig} and \ref{phia}, we show the results that we have obtained
considering $N=99$ and $M=21$ with the advanced approach in the direct
space, and we compare them with those achieved by means of a Fourier analysis.
In particular, in Fig.~\ref{eig}, we show the position on the Gauss plane
of all the obtained longitudinal wave vectors $\kappa_x$; and in
Fig.~\ref{phia}, we represent the behavior of the envelope function
$\Phi_{A \vec K}(y)$ corresponding to the eigenvalue with the largest real part.
As expected, the two methods yield exactly the same results.

Instead, in Fig.~\ref{eig1}, we show the values of the
longitudinal wave vectors $\kappa_x$ obtained with the simplified
treatment of the product term. As we can see from the highest order
wave vectors, this method is not equivalent to the Fourier one, even
though it yields quite good results for the low-order modes.

\section{Conclusion}
\label{conclusion}

We have proposed a method, based on a set of basis functions resulting from
the periodic repetition of sinc functions, for the numerical solution, in 
the direct space, of differential problems 
and, in particular, of quantum wave
equations, with periodic boundary conditions.

This approach is equivalent to a Fourier analysis and
(once we have chosen a given discretization) requires the
solution of a linear system with the same matrix size. It allows performing
a very efficient solution in the direct space, without
the need to Fourier transform the input functions and finally to inverse
transform the solutions.

We have applied this method to efficiently solve the
Dirac equation in an armchair graphene nanoribbon in the direct
space, verifying its equivalence to a reciprocal space approach.

We believe that this treatment also helps in clarifying the exact origin 
of the differences in the convergence behavior observed between 
direct space and reciprocal space approaches.

\appendix

\section{Orthonormality of the basis functions}
\label{orthonormality}

The orthonormality of two sinc functions centered on different sampling
points can be easily demonstrated in the following way:
\begin{eqnarray}
&&\displaystyle \int_{-\infty}^{+\infty}
\sinc\left(\frac{x-j\Delta}{\Delta}\right)
\sinc\left(\frac{x-\ell\Delta}{\Delta}\right)\,dx \nonumber\\[8pt]
&&\displaystyle =\left[\sinc\left(\frac{x}{\Delta}\right)\ast
\sinc\left(\frac{x}{\Delta}\right)\right]_{x=(\ell-j)\Delta} \nonumber\\[8pt]
&&\displaystyle =\left\{\mathcal{F}^{-1}
\left[\Delta^2\rect(\Delta f)\right]\right\}_{x=(\ell-j)\Delta}\nonumber\\[8pt]
&&\displaystyle =\Delta\,\sinc(\ell-j)=\Delta \,\delta_{j,\ell}\,,
\end{eqnarray}
where $\ast$ is the convolution operator and $\mathcal{F}^{-1}$ is the
inverse Fourier transform.

Moreover, if the function $z(x)$ is periodic with period $L$, we have that
\begin{eqnarray}
&&\displaystyle 
\frac{1}{\Delta}\int_0^L g_{j,\Delta}^*(\chi) z(\chi)\,d\chi=
\frac{1}{\Delta}\int_0^L g_{j,\Delta}(\chi) z(\chi)\,d\chi \nonumber\\
&&\displaystyle =\frac{1}{\Delta}\sum_{\eta=-\infty}^{+\infty}\int_0^L
\sinc\left(\frac{\chi-(j+\eta N)\Delta}{\Delta}\right)z(\chi)\,d\chi
\nonumber\\
&&\displaystyle 
=\frac{1}{\Delta}\sum_{\eta=-\infty}^{+\infty}\int_{-\eta L}^{(-\eta+1)L}
\sinc\left(\frac{\xi-j\Delta}{\Delta}\right)z(\xi+\eta N\Delta)\,d\xi
\nonumber\\
&&\displaystyle 
=\frac{1}{\Delta}\sum_{\eta=-\infty}^{+\infty}\int_{-\eta L}^{(-\eta+1)L}
\sinc\left(\frac{\xi-j\Delta}{\Delta}\right)z(\xi)\,d\xi \nonumber\\
&&\displaystyle =\frac{1}{\Delta}\int_{-\infty}^{+\infty}
\sinc\left(\frac{\xi-j\Delta}{\Delta}\right)z(\xi)\,d\xi\,,
\label{changedom}
\end{eqnarray}
where we have changed the integration variable from $\chi$ to
$\xi=\chi-\eta N\Delta=\chi-\eta L$.

From the previous observation, we derive the orthonormality of
the $g$ functions
\begin{eqnarray}
&&\displaystyle 
\frac{1}{\Delta}\int_0^L g_{j,\Delta}^*(x) g_{\ell,\Delta}(x)\,dx
\qquad\qquad\qquad\qquad\qquad\qquad\qquad \nonumber\\
&&\displaystyle 
=\frac{1}{\Delta}\int_{-\infty}^{+\infty}
\sinc\left(\frac{\xi-j\Delta}{\Delta}\right)g_{\ell,\Delta}(x)\,d\xi \nonumber\\
&&\displaystyle 
=\frac{1}{\Delta} \sum_{\eta=-\infty}^{+\infty}
\int_{-\infty}^{+\infty} \sinc\left(\frac{\xi-j\Delta}{\Delta}\right)
\sinc\left(\frac{x-(\ell+\eta N)\Delta}{\Delta}\right) \nonumber\\
&&\displaystyle 
=\frac{1}{\Delta}\sum_{\eta=-\infty}^{+\infty} \Delta \,\delta_{j,\ell+\eta N}
=\sum_{\eta=-\infty}^{+\infty} \delta_{j,\ell}\,\delta_{\eta,0}=\delta_{j,\ell}
\qquad\qquad\qquad
\end{eqnarray}
(with $j,\ell=0,\ldots,N-1$).

\section{Notes on the product term}
\label{productnotes}

Operating in the reciprocal space, if we substitute
\begin{equation}
\varphi(x)=\sum_{r=-D}^{D} a_r 
e^{i2\pi r \frac{\scriptstyle x}{\scriptstyle L}}\,,\quad
\tilde f(x)=\sum_{q=-Q}^{Q} f_q e^{i2\pi q \frac{\scriptstyle x}{\scriptstyle L}}
\end{equation}
(with $f_q$ being the DFT coefficients of $f(x)$ computed on its $N\,M=2\,Q+1$
samples) into the term $h(x)=\tilde f(x)\varphi(x)$ and then we consider
only the frequency components of the product within the interval $[-D/L,D/L]$,
we obtain that
\begin{equation}
h_0(x)=\underset{|q+r|\le D}{\sum_{q=-Q}^{Q}\,\sum_{r=-D}^{D}}
f_q a_r e^{i2\pi (q+r) \frac{\scriptstyle x}{\scriptstyle L}}
=\underset{|p-r|\le Q}{\sum_{p=-D}^{D}\,\sum_{r=-D}^{D}}
f_{p-r} a_r e^{i2\pi p \frac{\scriptstyle x}{\scriptstyle L}}\,.
\end{equation}
Since the functions $\exp(i2\pi p x/L)$ with $|p| \le D$ are band limited
with band $B\le 1/(2\Delta)$, they can be expressed in terms of their samples
taken with sampling interval $\Delta$
\begin{equation}
e^{i2\pi p \frac{\scriptstyle x}{\scriptstyle L}}=\sum_{\mu=0}^{N-1}
e^{i2\pi p \frac{\scriptstyle \mu}{\scriptstyle N}}\,
g_{\mu,\Delta}(x)\,.
\end{equation}
Moreover, exploiting Eq.~\eqref{coeff}, the coefficients $a_r$ and 
$f_{p-r}$ can be replaced with
\begin{eqnarray}
a_r&=&\frac{1}{N}\sum_{\ell=0}^{N-1}\varphi(\ell\,\Delta)\,
e^{-i2\pi r \frac{\scriptstyle \ell}{\scriptstyle N}}\,,\nonumber\\
f_{p-r}&=&\frac{1}{MN}\sum_{m=0}^{MN-1}f(m\,\Delta')\,
e^{-i2\pi (p-r)\frac{\scriptstyle m}{\scriptstyle MN}}\,.
\end{eqnarray}
After these substitutions, we obtain that
\begin{equation}
h_0(x)=\sum_{\mu=0}^{N-1}\left[\sum_{\ell=0}^{N-1}
\varphi(\ell\,\Delta)\beta_{\mu \ell}
\right]g_{\mu,\Delta}(x)
\label{h0bis}
\end{equation}
with
\begin{equation}
\beta_{\mu \ell}=
\frac{1}{MN^2}\sum_{m=0}^{MN-1}f\left(m\,\Delta'\right)\nu_{\mu \ell m}
\label{betabis}
\end{equation}
and
\begin{eqnarray}
\nu_{\mu \ell m}&=&
\displaystyle\underset{|p-r|\le Q}{\sum_{p=-D}^{D}\,\sum_{r=-D}^{D}}
\left[e^{i\frac{\scriptstyle 2\pi}{\scriptstyle N}
\left(\mu-\frac{\scriptstyle m}{\scriptstyle M}\right)}\right]^p
\left[e^{i\frac{\scriptstyle 2\pi}{\scriptstyle N}
\left(\ell-\frac{\scriptstyle m}{\scriptstyle M}\right)}\right]^{-r}
\nonumber\\
&=&\displaystyle\underset{|p+\tilde r|\le Q}
{\sum_{p=-D}^{D}\,\sum_{\tilde r=-D}^{D}}
\left[e^{i\frac{\scriptstyle 2\pi}{\scriptstyle N}
\left(\mu-\frac{\scriptstyle m}{\scriptstyle M}\right)}\right]^p
\left[e^{i\frac{\scriptstyle 2\pi}{\scriptstyle N}
\left(\ell-\frac{\scriptstyle m}{\scriptstyle M}\right)}
\right]^{\tilde r}
\label{nubis}
\end{eqnarray}
(with $\tilde r=-r$).

Equations~\eqref{h0bis} and \eqref{betabis} coincide with
Eqs.~\eqref{h0}--\eqref{beta} obtained operating in the direct space.
We can prove that Eq.~\eqref{nubis} coincides with
Eq.~\eqref{nu}, too. In detail, using Eq.~\eqref{trasfg}, we can write
the Fourier expansion of the $g$ functions as follows:
\begin{eqnarray}
& \displaystyle
g_{\mu,\Delta}(x)=\frac{1}{N}\sum_{p=-D}^{D}
e^{i2\pi p \frac{\scriptstyle x-\mu\Delta}{\scriptstyle L}}\,,&
\label{ser1}\\
& \displaystyle
g_{m,\Delta'}(x)=\frac{1}{MN}\sum_{q=-Q}^{Q}
e^{i2\pi q \frac{\scriptstyle x-m\Delta'}{\scriptstyle L}}\,,&
\label{ser2}\\
& \displaystyle
g_{\ell,\Delta}(x)=
\frac{1}{N}\sum_{r=-D}^{D}
e^{i2\pi r \frac{\scriptstyle x-\ell\Delta}{\scriptstyle L}}\,.&
\label{ser3}
\end{eqnarray}
Substituting Eqs.~\eqref{ser1}--\eqref{ser3} into Eq.~\eqref{nu},
we find that
\begin{eqnarray}
\nu_{\mu \ell m}&=&\displaystyle
\frac{1}{\Delta}\frac{1}{N}
\sum_{p=-D}^{D}\,\sum_{q=-Q}^{Q}\,\sum_{r=-D}^{D}
e^{i\frac{\scriptstyle 2\pi}{\scriptstyle N}
\left(p\mu-\frac{\scriptstyle q m}{\scriptstyle M}-r\ell\right)}
L \delta_{q,p-r}\nonumber\\
&=&\displaystyle
\underset{|p-r|\le Q}
{\sum_{p=-D}^{D}\,\sum_{r=-D}^{D}}
\left[e^{i\frac{\scriptstyle 2\pi}{\scriptstyle N}
\left(\mu-\frac{\scriptstyle m}{\scriptstyle M}\right)}\right]^p
\left[e^{i\frac{\scriptstyle 2\pi}{\scriptstyle N}
\left(\ell-\frac{\scriptstyle m}{\scriptstyle M}\right)}
\right]^{-r},
\end{eqnarray}
which coincides with Eq.~\eqref{nubis}.

If $M>1$, the condition $|p-r|\le Q$ (i.e., $|p+\tilde r|\le Q$)
is always satisfied and thus
\begin{equation}
\nu_{\mu \ell m}=
\sum_{p=-D}^{D}
\left[e^{i\frac{\scriptstyle 2\pi}{\scriptstyle N}
\left(\mu-\frac{\scriptstyle m}{\scriptstyle M}\right)}\right]^p
\sum_{\tilde r=-D}^{D}
\left[e^{i\frac{\scriptstyle 2\pi}{\scriptstyle N}
\left(\ell-\frac{\scriptstyle m}{\scriptstyle M}\right)}
\right]^{\tilde r}
=\lambda_1(m,\mu)\lambda_1(m,\ell)\,,
\end{equation}
which corresponds to Eq.~\eqref{mne1}.

Instead, if $M=1$, the condition $|p+\tilde r|\le Q$ has to be taken into
consideration. From Eq.~\eqref{nubis}, we see that for each
value of $p$, we have to sum over the integers $\tilde r$ for which
$-D \le \tilde r \le D$ and $-p-D \le \tilde r \le -p+D$
(note that $Q=D$ if $M=1$).
This set of values of $\tilde r$ corresponds for $p=0$ to
$-D \le \tilde r \le D$, for $p>0$ to $-D \le \tilde r \le -p+D$, and
for $p<0$ to $-p-D \le \tilde r \le D$. Therefore, Eq.~\eqref{nubis}
can be rewritten in the following way:
\begin{eqnarray}
\nu_{\mu \ell m}&=&\displaystyle
\sum_{\tilde r=-D}^{D}
\left[e^{i\frac{\scriptstyle 2\pi}{\scriptstyle N}(\ell-m)}\right]^{\tilde r}
\nonumber\\
&+&\displaystyle
\sum_{p=1}^{D}\Bigg\{
\left[e^{i\frac{\scriptstyle 2\pi}{\scriptstyle N}(\mu-m)}\right]^p
\sum_{\tilde r=-D}^{-p+D}
\left[e^{i\frac{\scriptstyle 2\pi}{\scriptstyle N}(\ell-m)}\right]^{\tilde r}
\Bigg\}\nonumber\\
&+&\displaystyle
\sum_{p=-D}^{-1}\Bigg\{
\left[e^{i\frac{\scriptstyle 2\pi}{\scriptstyle N}(\mu-m)}\right]^p
\sum_{\tilde r=-p-D}^{D}
\left[e^{i\frac{\scriptstyle 2\pi}{\scriptstyle N}(\ell-m)}\right]^{\tilde r}
\Bigg\}\,.\nonumber\\
\end{eqnarray}
Computing the first sum and noting that the third sum is the complex
conjugate of the second one, we can rewrite this expression as:
\begin{equation}
\nu_{\mu \ell m}=
N \delta_{m,\ell}+2\,{\rm Re}\Bigg\{\sum_{p=1}^{D}\bigg\{
\left[e^{i\frac{\scriptstyle 2\pi}{\scriptstyle N}(\mu-m)}\right]^p
\sum_{\tilde r=-D}^{-p+D}
\left[e^{i\frac{\scriptstyle 2\pi}{\scriptstyle N}(\ell-m)}\right]^{\tilde r}
\bigg\}\Bigg\}\,.
\end{equation}
If $\ell=m$, this is equal to
\begin{equation}
\nu_{\mu \ell m}=
N+2\,{\rm Re}\left\{\sum_{p=1}^{D}\left\{
\left[e^{i\frac{\scriptstyle 2\pi}{\scriptstyle N}(\mu-m)}\right]^p
(N-p)\right\}\right\}
=\lambda_2(m,\mu)\,.
\label{mne1a}
\end{equation}
If $\ell \ne m$ instead, we have that
\begin{eqnarray}
\displaystyle \nu_{\mu \ell m} &=& \displaystyle
2\,{\rm Re}\Bigg\{\sum_{p=1}^{D}\bigg\{
\left[e^{i\frac{\scriptstyle 2\pi}{\scriptstyle N}(\mu-m)}\right]^p
\frac{\displaystyle (-1)^{\ell-m}}
{\displaystyle 2 i \sin\left(\frac{\pi}{N}\,(\ell-m)\right)}
\left[e^{-i\frac{\scriptstyle 2\pi}{\scriptstyle N}p(\ell-m)}-1\right]
\bigg\}\Bigg\}\nonumber\\
&=& \displaystyle
\frac{\displaystyle (-1)^{\ell-m}}
{\displaystyle \sin\left(\frac{\pi}{N}\,(\ell-m)\right)}
\Bigg[{\rm Im}\bigg\{
\sum_{p=1}^{D} \left[e^{i\frac{\scriptstyle 2\pi}{\scriptstyle N}(\mu-\ell)}\right]^p
\bigg\}-{\rm Im}\bigg\{
\sum_{p=1}^{D} \left[e^{i\frac{\scriptstyle 2\pi}{\scriptstyle N}(\mu-m)}\right]^p
\bigg\}\Bigg]\nonumber\\
&=&\frac{\displaystyle (-1)^{\ell-m}\,
\left(\lambda_3(\ell,\mu)-\lambda_3(m,\mu)\right)}
{\displaystyle
\sin\left(\frac{\pi}{N}\,(\ell-m)\right)}\,.
\label{mne1b}
\end{eqnarray}
Equations~\eqref{mne1a} and \eqref{mne1b} correspond to Eq.~\eqref{meq1}.

\bibliographystyle{apsrev}

\end{document}